\documentclass{ifacconf}
\usepackage[usenames,dvipsnames]{xcolor}
\usepackage{graphicx}
\usepackage{natbib}
\usepackage{amssymb}
\usepackage{amsmath}
\usepackage{mathtools}
\usepackage{derivative}
\usepackage{threeparttable}
\usepackage{booktabs}
\usepackage{bm}
\usepackage[FIGTOPCAP]{subfigure}
\usepackage{float}
\newcommand{\tran}{{\mkern-1.5mu\mathsf{T}}}
\normalsize 
\graphicspath{{}}

\begin{document}
\begin{frontmatter}

\title{Enhancing reinforcement learning for population setpoint tracking in co-cultures}

\thanks[footnoteinfo]{This work was supported by the US DOE-BER (DE-SC0022155) and US NSF (MCB-2300239).}
\thanks[footnoteinfo]{\,SER e-mail: sebastian.espinelrios@csiro.au.}

\author[First,Fifth]{Sebastián Espinel-Ríos}
\author[Second]{Joyce Qiaoxi Mo}
\author[Third]{Dongda Zhang}
\author[Fourth]{Ehecatl Antonio del Rio-Chanona}
\author[First]{José L. Avalos}

\address[First]{Princeton University, United States}
\address[Second]{Princeton Satellite Systems, United States}
\address[Third]{University of Manchester, United Kingdom}
\address[Fourth]{Imperial College London, United Kingdom}
\address[Fifth]{Current affiliation: Commonwealth Scientific and Industrial Research Organisation, Australia}

\begin{abstract} 
Efficient multiple setpoint tracking can enable advanced biotechnological applications, such as maintaining desired population levels in co-cultures for optimal metabolic division of labor. In this study, we employ reinforcement learning as a control method for population setpoint tracking in co-cultures, focusing on policy-gradient techniques where the control policy is parameterized by neural networks. However, achieving accurate tracking across multiple setpoints is a significant challenge in reinforcement learning, as the agent must effectively balance the contributions of various setpoints to maximize the expected system performance. Traditional return functions, such as those based on a quadratic cost, often yield suboptimal performance due to their inability to efficiently guide the agent toward the simultaneous satisfaction of all setpoints. To overcome this, we propose a novel return function that rewards the simultaneous satisfaction of multiple setpoints and diminishes overall reward gains otherwise, accounting for both stage and terminal system performance. This return function includes parameters to fine-tune the desired smoothness and steepness of the learning process. We demonstrate our approach considering an \textit{Escherichia coli} co-culture in a chemostat with optogenetic control over amino acid synthesis pathways, leveraging auxotrophies to modulate growth.
\end{abstract}

\begin{keyword}
Reinforcement learning, policy gradient, return function, setpoint tracking, co-cultures, optogenetics.
\end{keyword}

\end{frontmatter}

\section{Introduction}
The genetic engineering of microorganisms has enabled the manufacturing of a wide range of products, including chemicals, fuels, materials, and pharmaceuticals \citep{nielsen_metabolic_2021}. Traditionally, bioproduction has relied on monocultures, harnessing the metabolic capabilities of a single species. However, engineering (large) metabolic pathways within a single cell can lead to \textit{metabolic burden}, compromising cell growth and the process volumetric productivity. To alleviate this burden and enhance the overall process efficiency, metabolic pathways can be partitioned among multiple microbial species or engineered strains in a consortium (\textit{division of labor}) \citep{roell_engineering_2019}. By distributing metabolic submodules across different populations, each member can specialize in the part of the pathway it is best suited for. Maintaining specific population levels within the consortium is essential for optimizing the production process, as the concentration of each specialized cell directly influences the achievable volumetric productivity of the metabolic submodule it carries.

Controlling population levels in microbial consortia, however, poses significant challenges. Due to the \textit{competitive exclusion principle} \citep{kneitel_gauses_2019}, when multiple microorganisms compete for a single limiting resource, e.g., the carbon source, the consortium member with the highest growth rate (\textit{fitness}) will eventually outcompete and displace the others. Adjusting initial inoculation ratios can \textit{alleviate} competitive exclusion in the \textit{short term}, but it does not alter the long-term dynamics, making this approach unsuitable for prolonged cultivations, such as those in continuous bioreactors. Engineering endogenous interactions, like mutualism, is another strategy to address competitive exclusion by creating co-dependency among cells, e.g., via engineering auxotrophies and cross-feeding relationships \citep{peng_molecular_2024}. However, the population dynamics are predetermined by the engineered interactions, limiting operational flexibility and adaptability.

To overcome these limitations, external control mechanisms have been proposed to enhance operational flexibility and facilitate feedback control via externally tunable inputs. This approach allows the user to define various population setpoints (i.e., constant reference values) according to process requirements, rather than relying on \textit{predefined engineered} setpoints as in endogenous interactions. Conventional control strategies, such as proportional-integral-derivative (PID) controllers, have been proposed to regulate population ratios in bioreactors \citep{gutierrez_mena_dynamic_2022}. Although simple and useful in certain cases, PID controllers are inherently \textit{reactive}, struggle with nonlinear system dynamics, and cannot handle system constraints, motivating more advanced control approaches. In the context of microbial consortia, model predictive control (MPC) offers a more advanced alternative by utilizing a system model to compute control actions that minimize a cost function \citep{espinel-rios_machine_2023}. However, its implementation may be challenging if obtaining accurate mathematical models is difficult or when dealing with complex models (e.g., stochastic, highly nonlinear, stiff, or discontinuous dynamics).

Reinforcement learning (RL) is a promising machine-learning control strategy where an agent (the \textit{controller}) learns optimal control actions (\textit{process inputs}) through interactions with the environment (the \textit{bioreactor system}). Previous studies have considered Q-learning, an action-value method, for setpoint tracking in microbial consortia with discrete bang-bang feeding control actions  \citep{treloar_deep_2020}. However, Q-learning involves \textit{deterministic} policies and requires careful balancing of exploration and exploitation (often \textit{hard-coded} via epsilon-greedy strategies). The \textit{value function} in Q-learning, the expected cumulative reward, from which the optimal actions are computed upon solving an optimization problem, can be approximated using, e.g., neural networks. Q-learning may struggle to converge to an optimal policy  if the value function is not properly approximated and/or if the optimization step poses numerical difficulties. This can be particularly challenging in continuous or high-dimensional action spaces. Thus, Q-learning may be more suitable for discrete actions.

In this work, we consider RL based on the policy-gradient method \citep{petsagkourakis_reinforcement_2020}, which can address several of the limitations of Q-learning. Policy-gradient methods directly optimize the \textit{control policy}, which can be approximated using, e.g., neural networks. This direct approach focused on the policy itself ensures convergence to at least a local optimum and naturally accommodates continuous input variables, enhancing operational flexibility as more of the input space can be explored and exploited. Furthermore, policy-gradient methods involve \textit{stochastic} policies, which naturally balance the agent's adaptive exploration-exploitation over time and are better suited for handling systems with high stochasticity, such as biological processes. Even when dealing with deterministic systems, a stochastic-by-design policy can favor exploration and help to escape local minima. Overall, RL offers a promising approach for controlling complex systems that are difficult to differentiate using conventional model-based optimization methods, while enabling the development of uncertainty-aware policies.

A critical component of RL is the design of the \textit{return function}, which guides the agent toward desired optimal behaviors. The inverse quadratic cost function, commonly used in optimal control problems, may offer good convergence properties when tracking a \textit{single} setpoint in RL. However, in the context of microbial consortia, the latter function lacks a mechanism to directly incentivize the simultaneous satisfaction of multiple independent setpoint objectives\footnote{By multiple \textit{setpoint} tracking, we refer to tracking \textit{constant} reference values for different state variables.}. In other words, the agent may need to explore more extensively to discover a \textit{sweet-spot} scenario where all setpoints are satisfied without prioritizing one over the other. This often results in a higher risk of suboptimal performance as the agent might oscillate between multiple \textit{individual} objectives. 

To tackle this issue, we propose a novel return function based on multiplicative inverse \textit{saturation} functions that can enhance multiple setpoint tracking performance. This reward structure ensures that maximum reward is achieved only when all setpoints are satisfied simultaneously \textit{and} that improving individual setpoints while others remain off-target \textit{diminishes} overall reward gains. This promotes a more balanced progression toward multiple targets and can guide the agent more precisely. The return function can be shaped to balance \textit{smoothness} and \textit{steepness} in the policy gradients and in the updates of its parameters. 

\section{Co-culture case study with optogenetic control of growth}
\label{sec:dynamic_model}
\begin{figure} [htb!]
\begin{center} 
\includegraphics[scale=0.36, trim={0 5 0 5}, clip]{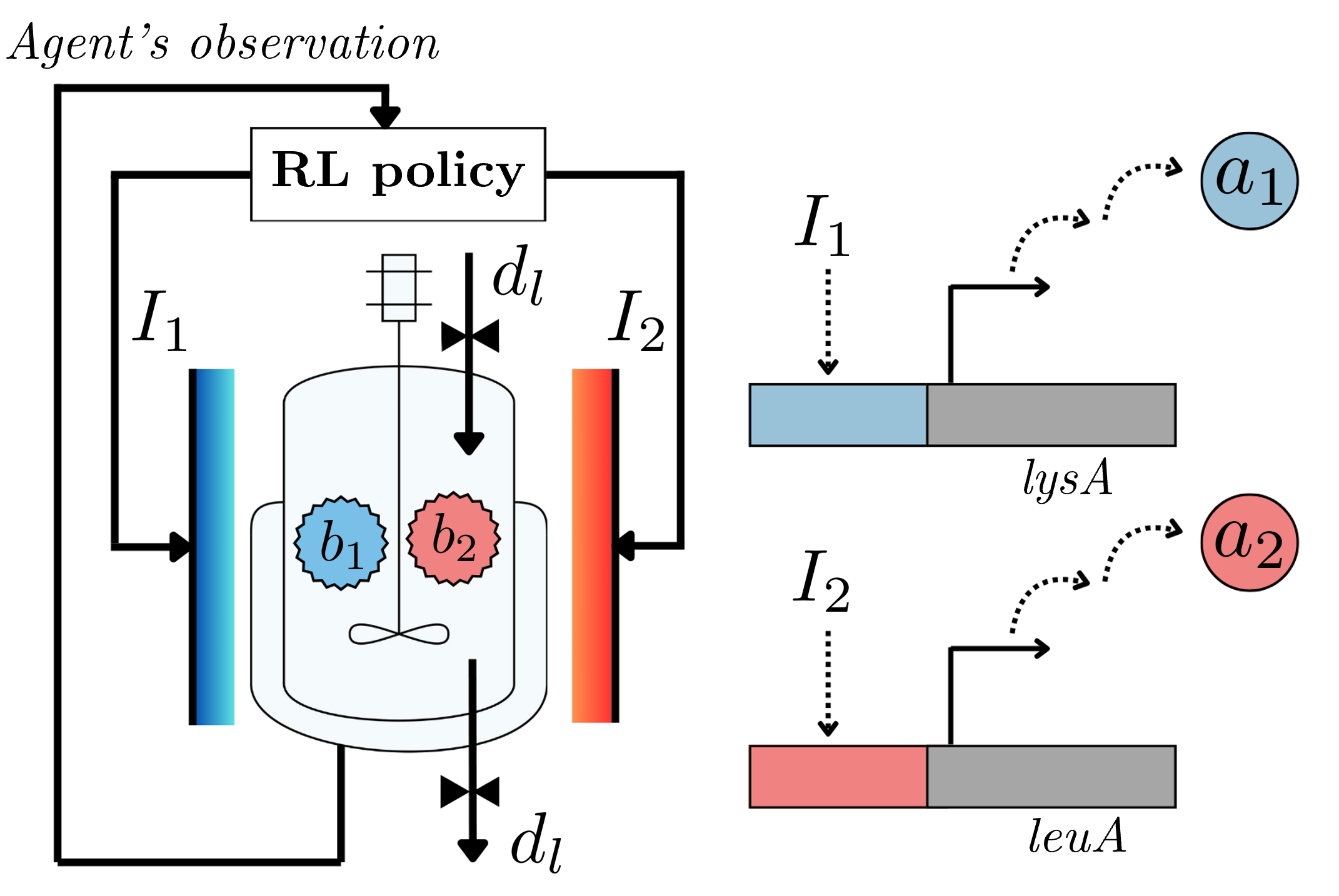}
\caption{\textit{E. coli} co-culture in a chemostat with optogenetic control of \textit{lysA} and \textit{leuA} synthesis.} 
\label{fig:schematic}
\end{center}
\end{figure}

As a case study, we consider a two-member \textit{Escherichia coli} consortium, where each microorganism has an auxotrophy for a specific amino acid (Fig. \ref{fig:schematic}). One strain, \textit{E. coli} 1, is auxotrophic for lysine due to the deletion of \textit{lysA} (diaminopimelate decarboxylase). The other strain, \textit{E. coli} 2, is auxotrophic for leucine due to the deletion of \textit{leuA} (2-isopropylmalate synthase). It is assumed that \textit{lysA} can be optogenetically induced with blue light in \textit{E. coli} 1, leveraging the PBLind-v1 system \citep{jayaraman_blue_2016}. Similarly, \textit{leuA} can be optogenetically induced with red light in \textit{E. coli} 2 using the pREDawn-DsRed system \citep{multamaki_optogenetic_2022}. For simplicity, we assume that the rate of auxotrophic amino acid synthesis can be directly linked to the light inputs, lumping the dynamics of enzyme expression (\textit{lysA} and \textit{leuA}). Additionally, we consider that amino acid synthesis does not lead to excretion, as amino acids accumulate only to normal physiological levels. The system dynamics thus follow:
\begin{subequations}
\begin{align}
    &\odv{s}{t} = - q_{s_1} b_1 - q_{s_2} b_2 + (s_{\text{in}} - s) d_l, \label{eq:s_ode}\\
    &\odv{b_i}{t} = (\mu_i - d_l) b_i, \quad \forall i\in\{1,2\}, \label{eq:b_i_ode}\\
    &\odv{a_i}{t} = q_{a_i} - (d_{a_i} + \mu_i) a_i, \quad \forall i\in\{1,2\} \label{eq:e_i_ode},
\end{align}
\end{subequations}
with growth, substrate uptake, and lumped transcription/translation kinetic rate functions:
\begin{subequations}
\begin{align}
    &\mu_i = \mu_{\text{max}_i} \left( \frac{s}{s + k_{s_i}} \right) \left( \frac{ f_{c} a_i}{ f_{c} a_i + k_{a_i}} \right), \quad \forall i\in \{1,2\}, \label{eq:mu_i}\\
    &q_{s,i} = Y_{s/b_i} \mu_i, \quad \forall i\in \{1,2\}, \label{eq:q_s_i}\\
    &q_{a,i} = q_{a_{\text{max}_i}} \left( \frac{I_i^{n_i}}{I_i^{n_i} + k_{I_i}^{n_i}} \right), \quad \forall i\in \{1,2\} \label{eq:q_e_i},
\end{align}
\end{subequations}
where the concentration of \textit{E. coli} $i$ in $\mathrm{g/L}$, the intracellular concentration of the amino acid counteracting the auxotrophy in species $ i $ in $\mathrm{mmol/g}$, and the concentration of the shared carbon source (glucose) in $\mathrm{mmol/L}$ are denoted by $ b_i \in \mathbb{R} $, $ a_i \in \mathbb{R} $, and $ s \in \mathbb{R} $, respectively. The control inputs are the blue and red light intensities, denoted by $ I_1 \in \mathbb{R} $ and $ I_2 \in \mathbb{R} $, respectively. Here, $ \mu_{\text{max}_i} $, $ k_{s_i} $, $ f_{c} $, $ k_{a_i} $, $ Y_{s/b_i} $, $ q_{a_{\text{max}_i}} $, $ n_i $, $k_{I_i}$, and $d_{a_i}$ are constant parameter values, $d_l$ is the dilution rate of the chemostat, and $s_{\text{in}}$ is the substrate concentration at the inflow.

\section{Reinforcement learning via policy gradients}
\label{sec:policy_gradient_rl}
Let us denote the dynamic states of the system by $\bm{x} \in \mathbb{R}^{n_x}$ and the system inputs by $\bm{u} \in \mathbb{R}^{n_u}$. The dynamic behavior of the system is formulated as a Markov decision process, where the transition from time $t$ to $t+1$ is described by a \textit{probability distribution} conditioned on the system state $\bm{x}_t$ and input $\bm{u}_t$. This transition can be approximated by the function $f_x: \mathbb{R}^{n_x} \times \mathbb{R}^{n_u} \rightarrow \mathbb{R}^{n_x}$, incorporating stochasticity through \textit{random noise} $\bm{d}_t \in \mathbb{R}^{n_x}$. Therefore, the system dynamics are given by:
\begin{equation}
    \bm{x}_{t+1} = f_x(\bm{x}_t, \bm{u}_t) + \bm{d}_t. \label{eq:system_dynamics}
\end{equation}
Note that the state vector $\bm{x}$ captures, in principle, all the information necessary to predict the dynamics of the system. Time is discretized in equidistant intervals, $t \in [t_0, t_1, ..., t_f]$.

The \textit{probability distribution} of the input actions is defined by the control policy $\pi$, parameterized by $\bm{\theta} \in \mathbb{R}^{n_\theta}$. Thus, the control action at time $t$ is sampled from:
\begin{equation}
    \bm{u}_{t} \sim \pi(\bm{u}_t \mid \boldsymbol{\bm{s}}_t, \bm{\theta}),
\end{equation}
where $\bm{s}_t \in \mathbb{R}^{n_s}$ represents the \textit{agent's observation} of the system, thus \textit{measurable}. Note that $\bm{s}_t$ is a more abstract variable that may not necessarily be the full \textit{dynamic system state} $\bm{x}_t$. That is, the agent may incorporate additional or different information to better understand the system when making decisions, such as past \textit{measured} state/input pairs or the process sampling time.

We represent with $\bm{\tau} \in \mathbb{R}^{n_\tau}$ the joint trajectory of \textit{observed} states, actions, and rewards $R \in \mathbb{R}$ over the time course of the process, i.e., $\bm{\tau} = \{ (\bm{s}_t, \bm{u}_t, R_{t+1}, \bm{s}_{t+1}) \}_{t=0}^{t_f-1}$, with probability distribution:
\begin{equation}
\mathrm{P}(\bm{\tau} \mid \bm{\theta}) =\mathrm{P}(\bm{s}_0) \prod_{t=0}^{t_f-1} \left[ \pi(\bm{u}_t \mid \bm{s}_t, \bm{\theta}) \mathrm{P}(\bm{x}_{t+1} \mid \bm{x}_t, \bm{u}_t) \right].
\label{eq:prob_dis_tau}
\end{equation}

We denote the agent's \textit{return function} or \textit{performance metric} with the function $J: \mathbb{R}^{n_\tau} \rightarrow \mathbb{R}$, which accounts for the rewards over $\bm{\tau}$. In RL, the agent aims to maximize the \textit{expected return} by finding an optimal policy:
\begin{equation}
\max_{\pi(\cdot)} \mathbb{E}_{\bm{\tau}} \left[ J(\bm{\tau}) \right],
\label{eq:rl_problem}
\end{equation}
where $\mathbb{E}_{\bm{\tau}}$ represents the expected value over $\bm{\tau}$ given the policy $\pi(\cdot)$.

We parameterize the mean $\bm{m}_t \in \mathbb{R}^{n_u}$ and standard deviation $\bm{\sigma}_t \in \mathbb{R}^{n_u}$ of the policy using deep neural networks $f_\mathrm{DNN}: \mathbb{R}^{n_s} \times \mathbb{R}^{n_\Theta} \rightarrow \mathbb{R}^{n_u} \times \mathbb{R}^{n_u}$ with parameters $\Theta \in \mathbb{R}^{n_\Theta}$:
\begin{equation}
\bm{m}_t, \bm{\sigma}_t = f_\mathrm{DNN} (\bm{s}_t, \bm{\Theta}).
\end{equation}
Thus, $\bm{\theta} := \bm{\Theta}$.

In policy-gradient RL, the parameters of the policy are updated following the gradient ascent: 
\begin{equation}
\bm{\theta}_{m+1} = \bm{\theta}_m + \alpha \nabla_{\bm{\theta}} \mathbb{E}_{\tau} \left[ J(\bm{\tau}) \right],
\label{eq:update_rule_general}
\end{equation}
where $m$ is an instance of parameter update, i.e., an \textit{epoch}, over  $N_{\mathrm{epoch}} \in \mathbb{N}$ epochs, and $\alpha \in \mathbb{R}$ is the learning rate. The first update occurs at $m=0$.

Based on the Policy Gradient Theorem \citep{NIPS1999_464d828b}:
\begin{equation}
\nabla_{\bm{\theta}} \mathbb{E}_{\tau} \left[ J(\bm{\tau}) \right] =  \int \mathrm{P}(\bm{\bm{\tau}} \mid \bm{\theta}) \nabla_{\bm{\theta}} \log \left( \mathrm{P}(\bm{\tau} \mid \bm{\theta}) \right) J(\bm{\tau}) \, \mathrm{d}\bm{\tau}, 
\label{eq:policy_gradient_theorem}
\end{equation}
thus,
\begin{equation}
\nabla_{\bm{\theta}} \mathbb{E}_{\bm{\tau}} \left[ J(\bm{\tau}) \right] =  \mathbb{E}_{\bm{\tau}} [ J(\bm{\tau}) \nabla_{\bm{\theta}} \log \left( \mathrm{P}(\bm{\tau} \mid \bm{\theta}) \right)].
\label{eq:sim_eq_gradient_exp}
\end{equation}

Combining Eqs. \eqref{eq:prob_dis_tau} and \eqref{eq:sim_eq_gradient_exp} leads to:
\begin{equation}
\nabla_{\bm{\theta}} \mathbb{E}_{\bm{\tau}} \left[ J(\bm{\tau}) \right] = \mathbb{E}_{\bm{\tau}} \left[ J(\bm{\tau}) \nabla_{\bm{\theta}} \sum_{t=0}^{t_f-1} \log \left( \pi\left( \bm{u}_t \mid \bm{s}_t, \bm{\theta} \right) \right)\right]. 
\label{eq:grad_exp}
\end{equation}

In this work, we approximate the expectation in Eq.~\eqref{eq:grad_exp} through Monte Carlo simulations:
\begin{align}
\nabla_{\bm{\theta}} \mathbb{E}_{\bm{\tau}} \left[ J\left( \bm{\tau} \right) \right]
&\approx \frac{1}{N_{\mathrm{MC}}} \sum_{k=1}^{N_{\mathrm{MC}}} \left[ \frac{J\left( \bm{\tau}^{(k)} \right) - \Bar{J}\left( \bm{\tau} \right)}{\sigma_J + \varepsilon} \right. \\
&\quad \times \nabla_{\bm{\theta}} \sum_{t=0}^{t_f - 1} \log\left( \pi\left( \bm{u}_t^{(k)} \mid \bm{s}_t^{(k)}, \bm{\theta} \right) \right) \bigg] \notag,
\label{eq:grad_exp_baseline}
\end{align}
where the \emph{episode} $k$ in $N_{\mathrm{MC}} \in \mathbb{N}$ Monte Carlo episodes is denoted by the superscript $(k)$. Note that we normalize the return by subtracting the mean return $\Bar{J}\left( \bm{\tau} \right)$ and dividing by the standard deviation of the return $\sigma_J$ across episodes, with a small constant $\varepsilon$ added for numerical stability. It is worth noting that the gradient of the log-probability informs how changing the parameters of the policy affects the probability of taking that action. Furthermore, values of $J\left( \bm{\tau}^{(k)} \right) > \Bar{J}\left( \bm{\tau} \right)$ will encourage updating the policy's parameters in a way that increases the probability of the actions taken in that episode, thereby reinforcing better-than-average returns. Conversely, values of $J\left( \bm{\tau}^{(k)} \right) < \Bar{J}\left( \bm{\tau} \right)$ will favor updating the policy's parameters to decrease the probability of those actions, thereby discouraging worse-than-average returns.

\section{Return function design}
\label{sec:return_design}
For setpoint tracking of the individual populations in the co-culture outlined in Section \ref{sec:dynamic_model}, we consider four possible return function designs.

\begin{itemize}
    \item \textbf{Case 1}. This design uses the stage quadratic objective as the return function:
    \begin{equation}
    J = - \sum_{t=1}^{t_f} \left[ w_1(b_{1_t} - b_{1}^*)^2 + w_2(b_{2_t} - b_{2}^*)^2 \right],
    \label{eq:inverse_quadratic}
    \end{equation}
    where $b_{1}^*$ and $b_{2}^*$ are the constant setpoint references for the respective biomass populations, and $w_1$ and $w_2$ are appropriate weights. Case 1 serves as a \textit{benchmark} return function. Eq. \eqref{eq:inverse_quadratic} in Eq. \eqref{eq:rl_problem} involves a \textit{weighted multi-objective optimization}, and there is no direct mechanism to favor the simultaneous satisfaction of both setpoints. This means that one can accumulate rewards even if only one setpoint improves, while the other does not, which can lead to stagnant learning. In that sense, if the weights are not \textit{tuned} properly, the agent may be biased or misled towards optimizing only one setpoint. Note that Eq. \eqref{eq:inverse_quadratic} deals with the \textit{inverse} quadratic cost, hence the negative sign. 

    \item \textbf{Case 2}. This design introduces multiplicative inverse \textit{saturation} functions in the return. The overall return is divided into a \textit{stage} reward accumulated from $t=1$ until $t=t_f-1$ and a \textit{terminal} or \textit{arrival} reward at the final time step $t= t_f$:
    \begin{equation}
    J = \sum_{t=1}^{t_f-1} \left[w_t q_{V_t} \right] + w_{t_f}q_{V_{t_f}},
    \label{eq:return_funtion_proposal}
    \end{equation}
    where for a given time $t \in \{t_1, ..., t_f\}$:
    \begin{equation}
    q_{V_{t}} = \beta_{V_{\text{max}}} \cdot \frac{\beta_{e_1}}{\beta_{e_1} + e_{1_t}} \cdot \frac{\beta_{e_2}}{\beta_{e_2} + e_{2_t}},
    \label{eq:saturation_multiplicative}
    \end{equation}
    and
    \begin{equation}
    e_{i_t} = (b_{i_t} - b_{i}^*)^2, \quad \forall i \in \{1,2\}.
    \end{equation}
    
    The maximum return at a given sampling time is determined by a tunable parameter $\beta_{V_{\text{max}}}$. The \textit{inverse} \textit{saturation} functions lead to a decrease in reward as the quadratic error of the biomass populations increases. The steepness of these functions is controlled by the tunable parameters $\beta_{e_1}$ and $\beta_{e_2}$, providing flexibility in how sharply the return improves with decreasing error. When the error $e_{i_t}$ approaches zero, the inverse saturation function associated with population $i$ approaches one and the maximum reward is achieved, i.e., $q_{V_{i_t}} = \beta_{V_{\text{max}}}$. This setup ensures that simultaneous setpoint satisfaction yields the highest reward. 
    
    Let us consider an extreme scenario, assuming $\beta_{e_1} = \beta_{e_2}$ for simplicity, where one biomass population is exactly at the target but the other one remains significantly off the target, then the overall stage reward gain will approach zero at that point. The incorporation of both \textit{stage} and \textit{terminal} rewards helps to balance transient and final multi-setpoint tracking performance, determined by the weights $w_t, \, \forall t\in \{1,...,t_f\}$. 
    
    In the context of the co-culture case study, we select $\beta_{e_1} = \beta_{e_2} = 3$ in Eq. \eqref{eq:saturation_multiplicative} for \textbf{Case 2}.

    \item \textbf{Case 3}. This design follows the same approach as Case 2, but with $\beta_{e_1} = \beta_{e_2} = 9$ in Eq. \eqref{eq:saturation_multiplicative}.
    
    \item \textbf{Case 4}. This design also follows the approach of Case 2, but with $\beta_{e_1} = \beta_{e_2} = 27$ in Eq. \eqref{eq:saturation_multiplicative}.
\end{itemize}

\begin{figure} [htb!]
\begin{center} 
\includegraphics[scale=0.5, trim={0 12 0 1}, clip]{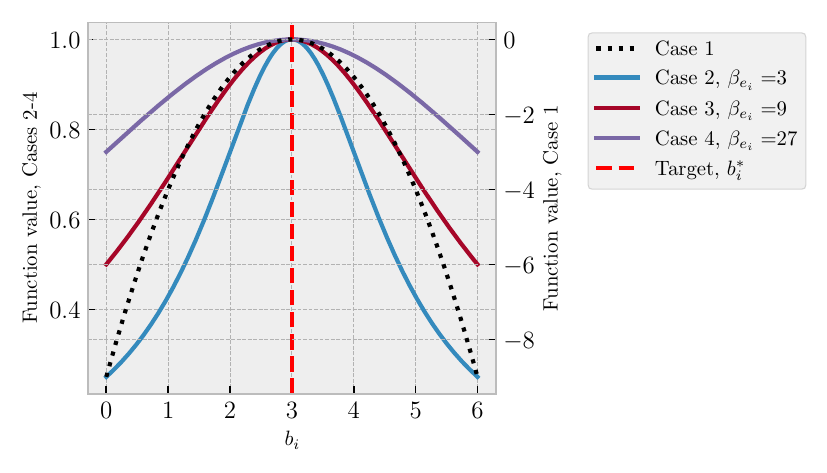}
\caption{Comparison of key individual elements contributing to the return functions in Cases 1-4. For Case 1, the inverse quadratic error term \mbox{$-(b_{i_t} - b_{i}^*)^2$} is plotted, representing the penalty for deviations from the target. For Cases 2-4, the inverse saturation function \mbox{$\frac{\beta_{e_i}}{\beta_{e_i} + e_{i_t}}$} is shown. The target for population $i$ is $b_{i}^*=3 \, \mathrm{g/L}$.}
\label{fig:functions_return}
\end{center}
\end{figure}

The key individual functions contributing to the returns for Cases 1-4 are shown in Fig. \ref{fig:functions_return}. As seen in the figure, the functions exhibit varying degrees of steepness and smoothness depending on the distance to the target.

\section{Setpoint tracking results in the co-culture}
\label{sec:results}

\begin{figure*}[h!]
	\begin{center}    
		\subfigure[Case 1]{\includegraphics[scale=0.43, trim={0 10 0 10}, clip]{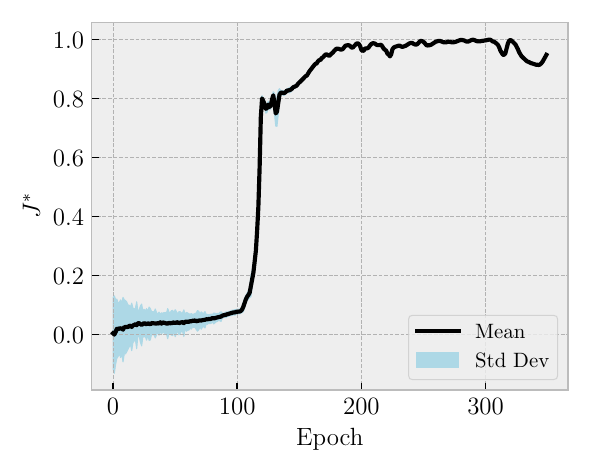}}
		\subfigure[Case 2]{\includegraphics[scale=0.43, trim={0 10 0 10}, clip]{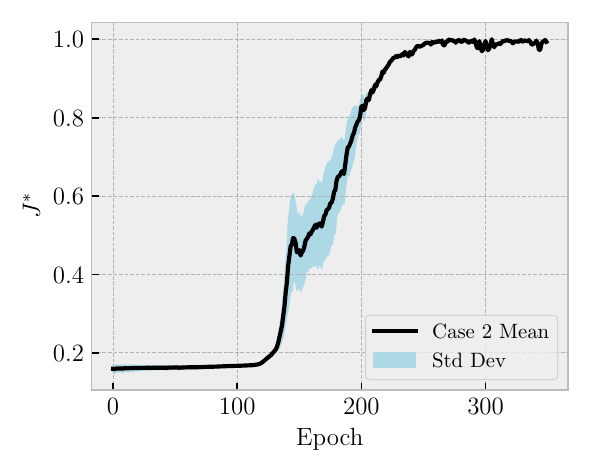}}
		\subfigure[Case 3]{\includegraphics[scale=0.43, trim={0 10 0 10}, clip]{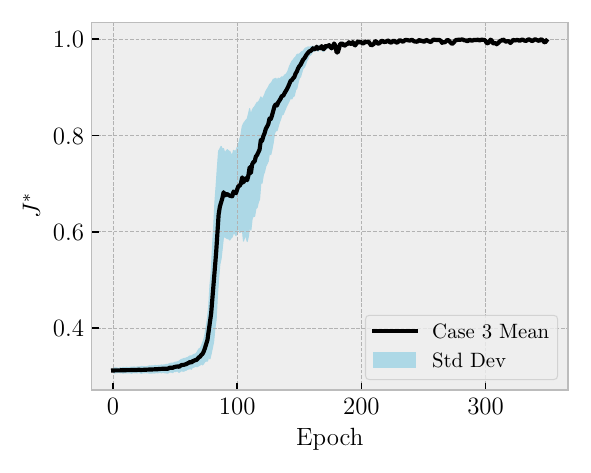}}
		\subfigure[Case 4]{\includegraphics[scale=0.43, trim={0 10 0 10}, clip]{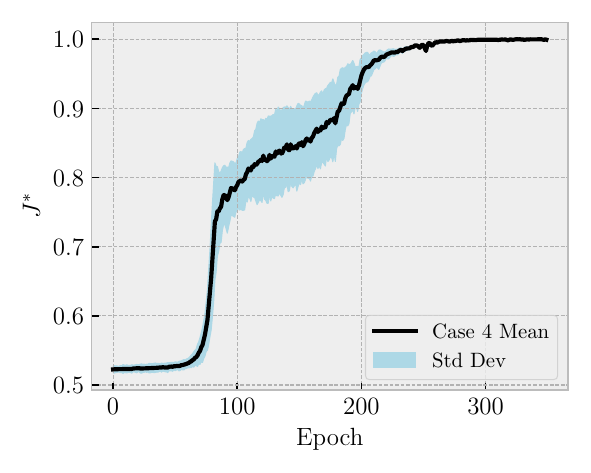}}
		\caption{Normalized return function $J^*$ over 350 epochs for Cases 1-4. Both the mean and standard deviation (Std Dev) are shown.}
		\label{fig:return_function}
	\end{center}
\end{figure*}

\begin{figure}[h!]
	\begin{center}    
		\subfigure[Case 1, \textit{E. coli} 1]{\includegraphics[scale=0.43, trim={0 5 0 10}, clip]{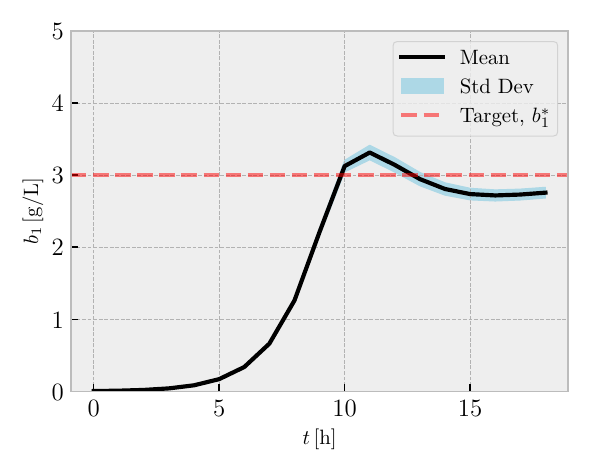}}
		\vspace{-0.3cm}
		\subfigure[Case 1, \textit{E. coli} 2]{\includegraphics[scale=0.43, trim={0 5 0 10}, clip]{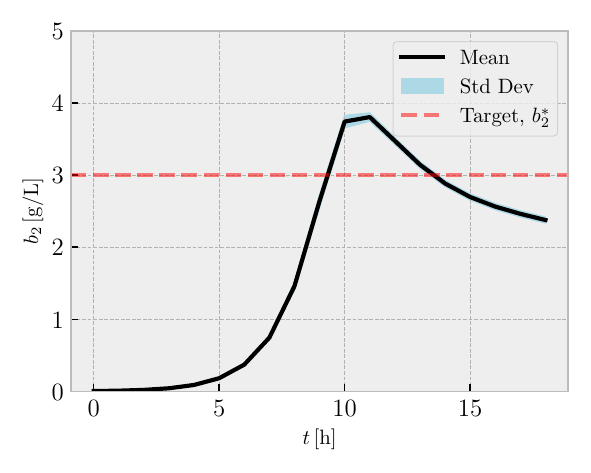}}
		\vspace{-0.3cm}
		\subfigure[Case 2, \textit{E. coli} 1]{\includegraphics[scale=0.43, trim={0 5 0 10}, clip]{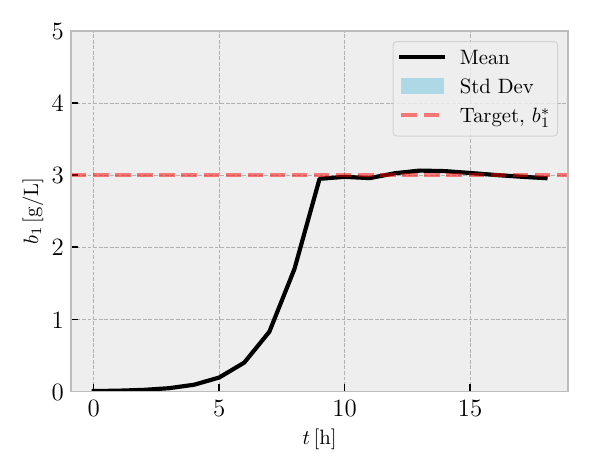}}
		\subfigure[Case 2, \textit{E. coli} 2]{\includegraphics[scale=0.43, trim={0 5 0 10}, clip]{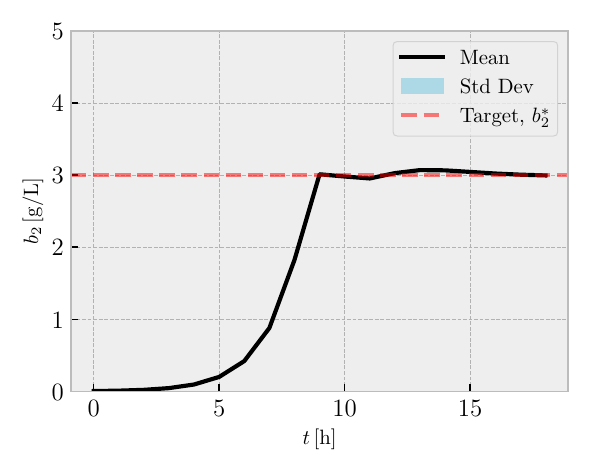}}
		\vspace{-0.3cm}
		\subfigure[Case 3, \textit{E. coli} 1]{\includegraphics[scale=0.43, trim={0 5 0 10}, clip]{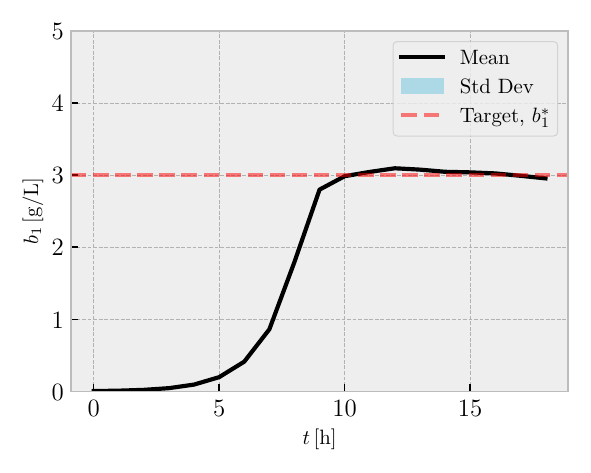}}
		\subfigure[Case 3, \textit{E. coli} 2]{\includegraphics[scale=0.43, trim={0 5 0 10}, clip]{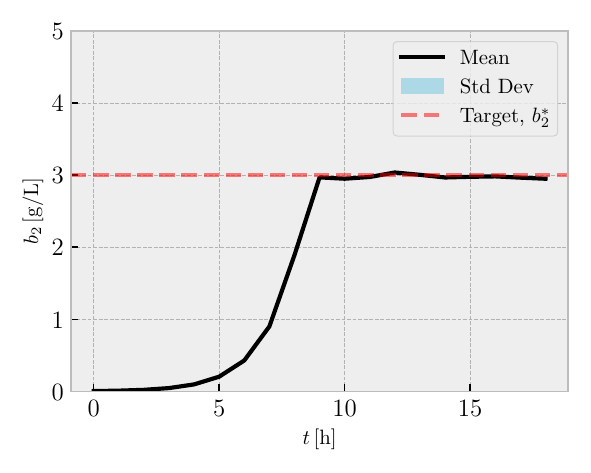}}
		\vspace{-0.3cm}
		\subfigure[Case 4, \textit{E. coli} 1]{\includegraphics[scale=0.43, trim={0 5 0 10}, clip]{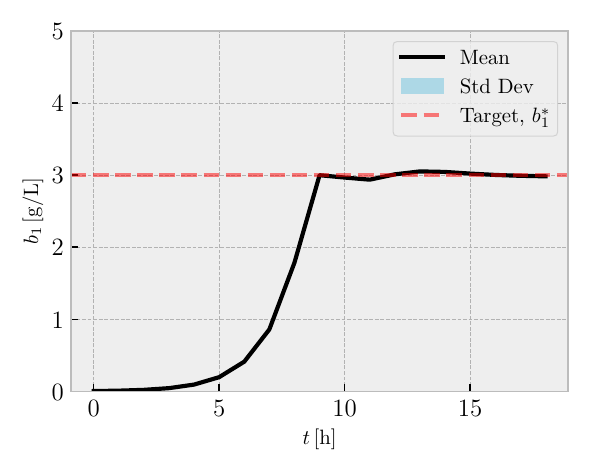}}
		\subfigure[Case 4, \textit{E. coli} 2]{\includegraphics[scale=0.43, trim={0 5 0 10}, clip]{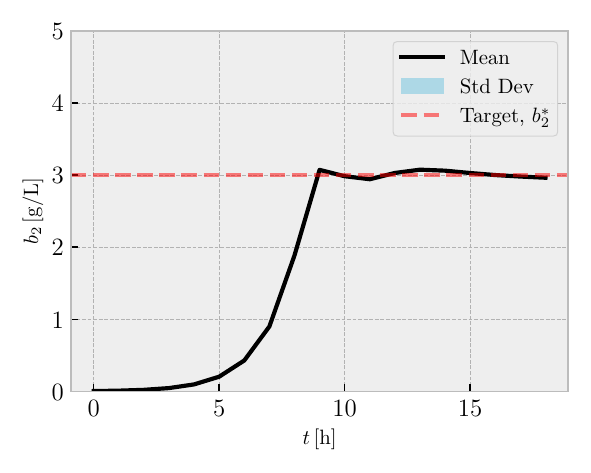}}
		\caption{Setpoint tracking performance of population levels in the co-culture for Cases 1-4. Both the mean and the standard deviation (Std Dev) are shown. These scenarios correspond to the epochs with the highest return function for each scenario (cf. Figure \ref{fig:return_function}).}
		\label{fig:setpoint_tracking}
	\end{center}
\end{figure}

We compare the performance of the return functions described in the previous section for multi-setpoint tracking of populations in our case study using RL. The setpoints are $b_{1}^* = b_{2}^* = 3 \, \mathrm{g/L}$. The learning process is conducted in PyTorch \citep{pytorch} over 350 epochs, 500 Monte Carlo episodes each, with a constant learning rate of 0.001. The policy is parameterized using a deep neural network consisting of 4 hidden layers with 20 nodes each, employing the LeakyReLU activation function. There are two output linear layers: one producing the means and the other producing the standard deviations of the inputs. The dynamic model in Eqs. \eqref{eq:s_ode}-\eqref{eq:q_e_i} is used to simulate the chemostat bioreactor in CasADi \citep{Andersson2019}. We assume full state observability and use two past state/input pairs and a time embedding normalized in the range $[-1,1]$ as the agent's observations, resulting in the vector $s_t := [\bm{x}_{t-1}^\tran, \bm{u}_{t-2}^\tran, \bm{x}_{t}^\tran, \bm{u}_{t-1}^\tran, t_t^*]^\tran$, where $t_t^*$ represents the time embedding. Given our focus on return function design, we assume a deterministic plant for simplicity; yet we maintain a stochastic policy to promote \textit{natural} exploration and avoid local minima. The simulation is carried out over 18 equally spaced time steps of 1 hour each, i.e., $t_f = 18 \, \mathrm{h}$. For Cases 2-3, $\beta_{\text{max}}=1$, and the weights $w_1 = w_2 = ... = w_{t_f-1} = 1$. In addition, we set $w_{t_f} = 2$ to promote stability toward the endpoint of the process. 

The \textit{initial} values for these states were set to $s = 5.5 \, \mathrm{mmol/L}$, $b_1 = b_2 = 0.005 \, \mathrm{g/L}$, $a_1 = 1.545\times
10^{-2} \, \mathrm{mmol/g}$, and $a_2 = 1.655\times 10^{-3}
 \, \mathrm{mmol/g}$. The model parameters are $\mu_{\text{max}_1} = \mu_{\text{max}_2} = 0.982 \, \mathrm{h^{-1}}$, $k_{s_1}=k_{s_2} = 2.964\times10^{-4} \,  \mathrm{mmol/L}$, $f_{c} = 1100 \,  \mathrm{g/L}$, $k_{a_1} = 1.7 \, \mathrm{mmol/L}$, $k_{a_2} = 0.182 \, \mathrm{mmol/L}$, $Y_{s/b_1}=Y_{s/b_2} = 10.18 \, \mathrm{mmol/g}$, $q_{a_{\text{max}_1}} = 0.337 \, \mathrm{mmol/g/h}$, $q_{a_{\text{max}_2}} = 0.036 \, \mathrm{mmol/g/h}$, $n_1 = 2$, $k_{I_1} = 1.052 \, \mathrm{W/{m^2}}$, $n_2 =  4.865$, $k_{I_2} = 1.34 \, \mathrm{\mu W/{cm}^2}$, $d_l = 0.15 \, h^{-1}$, $s_{\text{in}} = 200 \, \mathrm{mmol/L}$.

Fig. \ref{fig:return_function} shows the evolution of the return function over 350 epochs for Cases 1–4. The biomass profiles for the corresponding best-performing epochs (i.e., yielding the highest mean return values) are presented in Fig. \ref{fig:setpoint_tracking}. Case 1 shows a very steep improvement in the return function shortly after 100 epochs, then slows down until stagnating around 200 epochs, with slight oscillations thereafter. As expected, the setpoints were not properly reached at the \textit{best} epoch for Case 1. The system initially overshoots the target, then undershoots it, and does not stabilize at the desired references over the considered epochs. In contrast, the proposed return function manages to stabilize the biomass populations at the desired setpoint references in all remaining cases, i.e., Cases 2-4. This is due to the emphasis on the simultaneous satisfaction of the setpoints, where the overall reward is diminished if any state is off-target. The main difference among these scenarios is how fast the system converges to the setpoint references. For example, Cases 2-4 plateau around 250, 180, and 250 epochs, respectively.

The high smoothness of the return function in Case 4 (cf. Fig. \ref{fig:functions_return}) makes the agent experience less \textit{aggressive} parameter updates (cf. Eq. 
\eqref{eq:update_rule_general}), explaining its slow convergence.  This scenario offers a \textit{safe} strategy when smooth convergence and learning stability are prioritized, despite the larger number of epochs required. In contrast, Case 3 finds the \textit{best} balance between steepness and smoothness in the return function, leading to the fastest convergence. Notably, although Case 2 has the steepest shape (cf. Fig. \ref{fig:functions_return}), it takes about the same number of epochs as Case 4 to stabilize (the latter having the smoothest return function shape). This can be explained by the fact that, similar to Case 1, high steepness can lead to more aggressive parameter updates and, consequently, less efficient learning.

Overall, these results underscore the importance of appropriate return function design for efficient RL, particularly in policy-gradient approaches and in multiple setpoint tracking problems. One advantage of our proposed return function design is that it can be \textit{shaped} by the user to achieve either steeper or smoother convergence. Furthermore, this approach can unlock schemes with online adaptation of the proposed return function’s parameters to better guide the agent’s learning. Such an adaptive strategy would play a similar role to an \textit{adaptive learning rate} in Eq. \eqref{eq:update_rule_general}, but with the advantage of being directly tailored to the return function, making it more \textit{interpretable}.

\begin{figure*}[h!]
    \begin{center}    
        \subfigure[Input for \textit{E. coli} 1]{\includegraphics[scale=0.43, trim={0 10 0 10}, clip]{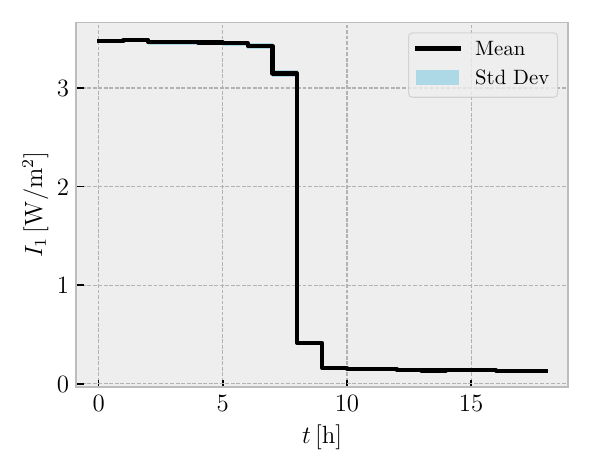}}
        \subfigure[Lysine \textit{E. coli} 1]{\includegraphics[scale=0.43, trim={0 10 0 10}, clip]{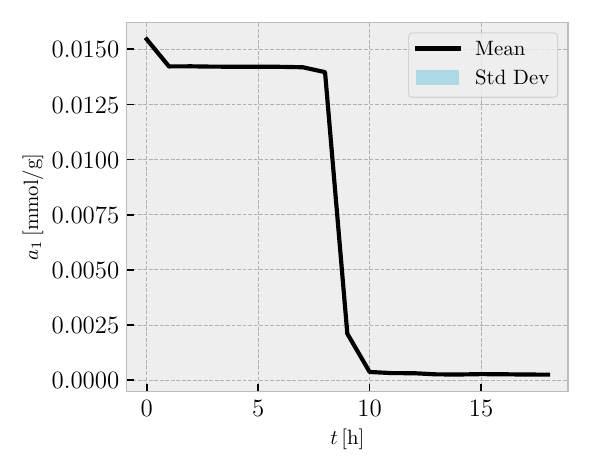}}
        \subfigure[Input for \textit{E. coli} 2]{\includegraphics[scale=0.43, trim={0 10 0 10}, clip]{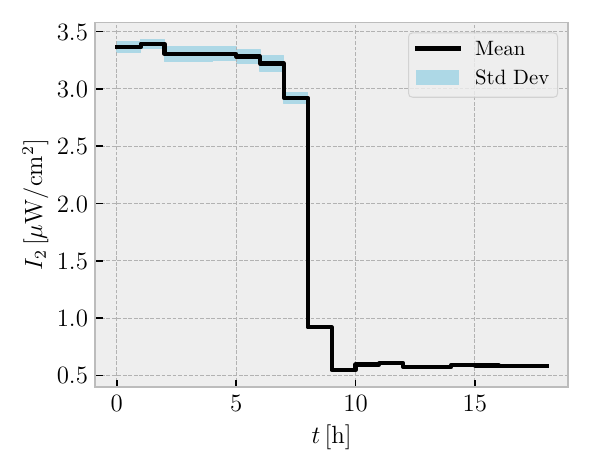}}
        \subfigure[Leucine in \textit{E. coli} 2]{\includegraphics[scale=0.43, trim={0 10 0 10}, clip]{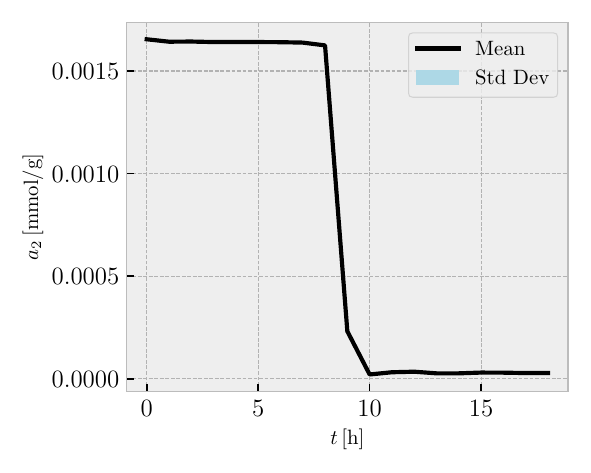}}
        \caption{Input and amino acid trajectories for the epoch with the highest mean return function value in Case 3.}
        \label{fig:input_amino_acids}
    \end{center}
\end{figure*}

\begin{figure} [h!]
\begin{center} 
\includegraphics[scale=0.49, trim={0 12 0 1}, clip]{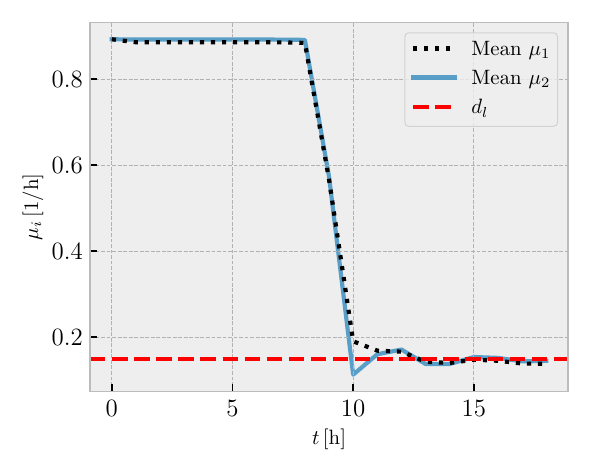}
\caption{Mean growth rates for the epoch with the highest mean return function value in Case 3.} 
\label{fig:growth_rates_dl}
\end{center}
\end{figure}

Finally, for demonstration purposes, we show in Fig. \ref{fig:input_amino_acids} the dynamic profiles of the inputs and intracellular amino acids for which the microorganisms are auxotrophic. As expected, amino acid accumulation rates correlate with light intensities. Also, as observed in Fig. \ref{fig:growth_rates_dl}, the agent successfully maintains the system at the setpoint references by regulating the growth rates to match the bioreactor’s dilution rate once the desired biomass concentrations are reached. In other words, the agent learns to influence the system’s transient dynamics to drive it toward the desired \textit{steady state}, where the rate of new cell generation balances with the rate of cell removal.

It is worth noting that the presented RL strategy is, in principle, model-free. However, dynamic models, even if only approximations of the real system, can be used to pretrain the policy offline before interaction with the real system. In addition, \textit{domain randomization} during training, i.e., augmenting the system with \textit{a priori} \textit{known} or \textit{expected} uncertainties, such as disturbances, stochastic dynamics, and variable initial conditions, can be adopted to improve the robustness of the control policy. Furthermore, offline RL strategies can be used to derive policies from \textit{process data} without active interaction with the environment, or to perform \textit{behavioral cloning} based on available \textit{expert} policies.

\section{Conclusion}
\label{sec:conclusion}
In this work, we presented a novel return function design for policy-gradient RL tailored for setpoint tracking of multiple targets, such as in the case of population control in co-cultures. The proposed return function explicitly rewards the simultaneous satisfaction of multiple setpoints, leading to improved performance compared to the standard quadratic cost function, which served as a benchmark. Moreover, it can be tuned by adjusting appropriate parameters, enabling control over the smoothness and steepness of the learning process. The outlined approach can facilitate the development and application of RL for microbial consortia in biotechnological production. Future work will assess the robustness of this method under uncertainty, particularly in scenarios involving multiple setpoint and trajectory tracking (time-varying references) in microbial consortia. Another promising direction is the adaptive tuning of the return function’s parameters to enhance learning efficiency and convergence.

\bibliography{references}

\end{document}